# Can the Stark-Einstein law resolve the measurement problem from an *animate* perspective?

Fred H. Thaheld *

4000 Alan Shepard St. #256, Sacramento, CA 95834, USA

*Nobody knows what quantum mechanics says exactly about any situation, for nobody knows where the boundary really is between wavy quantum systems and the world of particular events.*

<div style="text-align: right">John Bell</div>

**Abstract**

   Analysis of the Stark-Einstein law as it applies to the retinal molecule, which is part of the rhodopsin molecule within the rod cells of the retina, reveals that it may provide the solution to the measurement problem from an *animate* perspective. That it represents a natural boundary where the Schrödinger equation or wave function automatically goes from linear to nonlinear while remaining in a deterministic state.  It will be possible in the near future to subject this theory to empirical tests as has been previously proposed.  This analysis provides a contrast to the many decades well studied and debated *inanimate* measurement problem and would represent an addition to the Stark-Einstein law involving information carried by the photon.

1. Introduction

   The measurement problem in quantum mechanics deals with the issue of how or whether wave function collapse occurs when a physical quantity is measured and what role measurements play in quantum mechanics (Wigner, 1963; Leggett, 2005; Adler and Bassi, 2009). How does the wave function go from linear to nonlinear?  The wave function evolves continuously according to the Schrödinger equation as a linear superposition of different states, but actual measurements always find the physical system in a definite state with regards to

Tel: +1 916 928 4461 *E-mail address:* fthaheld@directcon.net

physical quantities such as photon polarization or electron spin. We cannot predict precise results for measurements of this nature, only probabilities. Any future evolution is based on the state the system was discovered to be in when the measurement was made, so the measurement "did something" to the system that is not obviously a consequence of the Schrödinger evolution.

There appears to be a major difference between *inanimate* and *animate* "measurements", which is critical to the issue of any discussion of the measurement problem from this perspective. In the *inanimate* instance we have been mainly interested in measuring superposed photon polarization or superposed electron spin states, where the total of the information possessed by the photons concerns polarization states, and for electrons their spin states. As is well known, the outcome probabilities are given by the absolute value squared of the corresponding coefficient in the initial wave function, or the Born 50-50 rule, and therefore the outcomes are not predictable in advance. Since we prepare these superimposed states, they only contain the information which we are interested in deriving when we set up the experiment. I.e., we are only interested in very limited polarization or spin outcome information, especially as regards the measurement problem.

Since this proposal will be dealing only with photons, it is important to point out that the photon is a fundamental carrier of information, possessing numerous information carrying degrees of freedom in addition to polarization. These include frequency, phase, arrival time, orbital angular momentum, linear momentum, entanglement, etc. This can be considered as *intrinsic* or *inherent* quantum information possessed by a photon, in contrast to the additional *extrinsic* classical information which can be acquired by the photon from the natural classical *environment* with which we are all familiar from a visual perspective, and which can possess an infinite number of definite visual possibilities. For example, a photon that has been emitted or scattered by the text projected on a computer screen or printed on a sheet of paper, carries information of this text at the quantum level and an observer acquires this information by intercepting a small fraction of these photons (Zurek, 2007). This classical information has been reduced down to the quantum level by the photon(s) and represents an exact copy or copies with accompanying wave functions.

**2. State preparation prior to the operation of the Stark-Einstein law**

In this *animate* measurement instance, these photons have either naturally scattered off of a multitude of different surfaces or been naturally emitted by various sources (mostly without our specific input), and they therefore contain *classical environmental information* which is constantly being presented to the retina (Zurek, 2007). In this specific visual case, the preparation of these states by Nature means that the outcome probability results in the *naturalization* of the Born rule (as was previously pointed out), so that it is no longer a 50-50 result but, 100% (Thaheld, 2009)! (It has been recently brought to my attention that this is the equivalent of saying "probability 1", as encompassed by the deterministic tenets of QBism, a topic which will be discussed later in this paper). And, since Nature prepares these states (again, mostly without our specific input), they can be considered as a true picture or representation of *classical environmental information* as expressed in their wave functions and can therefore be regarded as *pure* states representing maximal knowledge about these states and their preparation (Chiribella, D'Ariano, Perinotti, 2011; Hardy, 2001). There is also the possibility that we might be dealing with *mixed* states which are part of a larger *pure* state, in which case it would still be possible to describe each physical process with maximum information. This is also known as the "*purification principle"* (Chiribella, D'Ariano, Perinotti, 2010).

Now, with this picture in mind, exactly what would constitute a measurement and, at what point might superpositions break down and definite outcomes appear in an *animate* visual setting? At this point we bring in the Stark-Einstein law.

**3. Stark-Einstein law and retinal molecule**

The Stark-Einstein law is named after Johannes Stark and Albert Einstein, who independently formulated the law between 1908 and 1913 (Cox and Kemp, 1971). It is also known as the photochemical equivalence law or photoequivalence law. In essence it says that each quantum of light that is absorbed by a molecule will cause a (primary) chemical or physical reaction in that molecule. And, although it was first proposed for physics and chemistry in the inorganic material world, it has a great potential in the field of biology as outlined herein from an *information* perspective.

It is important to stress here, for the first time to the author's knowledge, that in addition to the chemical or physical reaction mentioned, that *classical information* acquired from the *environment* by the photons, and thereby reduced to the quantum level, will also be passed on, not only to all the inorganic and organic molecules but, especially to the very receptive retinal molecules, to be further utilized, based upon the quantum detection efficiency of the retinal molecules. The retinal molecule is in an extremely unique position with regards to nearly all the other numerous organic and inorganic molecules in this regard, in being able to utilize this *classical environmental information* rather than it being discarded and lost for all time. This is the concept which the author feels should be a natural addition to the S-E law.

Let us now have a photon be absorbed by retinal, which is a light sensitive molecule found in the photoreceptor cells of the retina. Retinal $C_{20}H_{28}O$ is the fundamental chromophore involved in the transduction of light into electrical signals, which are processed by other cells in the retina and then sent to the brain where they produce visual images (Baylor, 1996; Rieke and Baylor, 1998; Whikehart, 2003).

To briefly recaptulate, there are ~$10^8$ rod cells in each human eye or retina, with ~$10^8$ rhodopsin molecules in each rod cell and, with each rhodopsin molecule containing a retinal molecule (Whikehart, 2003). Rod cells are natural photodetectors and represent a natural biological interface with photons. They convert incident light into electrical signals, which are then sent to the brain via the optic nerve. It is critical to this analysis to mention that all the rhodopsin molecules are *identical*, as are all the retinal molecules. The rod cell absorbs photons with a quantum detection efficiency of 29$\pm$ 4.7%, and absorbed photons produce detectable output signals (Baylor, 1996; Rieke and Baylor, 1998; Phan et al, 2013). This means that only 1 out of the ~$10^8$ rhodopsin molecules within each rod cell, and its embedded retinal molecule, is involved sequentially each time in one successful absorption event which encompasses photoexcitation, photoisoimerization and phototransduction. Simultaneously, exposed to a continuous photon stream from the environment, the other ~$10^8$ rod cells are undergoing this same process, subject to the quantum detection efficiency. The only probabilistic question, which is of no importance to us in this analysis, is which one of the identical ~$10^8$ retinal molecules in each rod cell, will end up successfully absorbing a photon each time. Once a retinal molecule absorbs a photon a lengthy process begins, culminating in an amplified current ~1 pA in amplitude and lasting ~200 ms, resulting in 2-3 signals in the rod cell's synaptic junction, which eventually leads into an axon and from there to the optic nerve. (Baylor, 1996; Rieke and Baylor, 1998; Whikehart, 2003). I.e., the initial quantum *environmental information* has been amplified back to the classical level in a reversible fashion. In addition, the discs within the rod cell which contain the retinal and rhodopsin molecules, along with these molecules, are continuously being shed and generated in a cyclical fashion (Mazzolini et al, 2015).

Prior to successful absorption, which ultimately constitutes phototransduction and a detectable output signal, a photon can be considered to be either a wave or a particle but, it has to ultimately be a particle in order to be absorbed by this molecule, in line with the S-E law. When a photon is absorbed by the retinal molecule into one of the $\pi$ bonds found between carbon 11 and 12, it passes on its energy, and more importantly its *information* simultaneously, to an electron in the highest $\pi$ orbital, which then jumps into a higher $\pi^*$ electron orbit (Thaheld, 2008). One now has to ask the question as to whether the wave function collapsed at the instant of absorption when the photon interacted with the $\pi$ electron and passed on its energy and *information*? In any case, whenever and wherever this collapse or absorption takes place, the outcome will be the same each time after repeated measurements, and it will be governed by the S-E law, which acts as a boundary between linear and nonlinear states. And, this *information* initially derived from the *classical environment*, will pass from an *inanimate* state to an *animate* state at this point in time.

We now go from Schrödinger linear deterministic superposed states, which can be *pure* states or *mixed* states as part of a *pure* state to a Schrödinger nonlinear deterministic collapsed *pure* state, both of which states possess the same *information* every time but, in different amounts (Chiribella, D'Ariano, Perinotti, 2011)! It is just that the superposed states possess more of this same *information*, while the collapsed *pure* state is a single copy of this same *information*. What I am saying is that both states contain the same information derived from the *classical environment* so that there is never any guessing as to what the final collapse outcome will be. I base this on the fact that none of us ever observe 2 different versions or superpositions of an identical *classical environment*. That the photon acquires *classical information* from the *environment* which, while it is reduced down to the quantum level by the photon, still maintains its inherent classicality, waiting to be amplified back to the classical level. Since Nature constantly prepares these classical *extrinsic* states in which all properties of the system are fixed and certain, instant by instant as related to the velocity of light, they are in a *pure* state with probability 1 (Chiribella, D'Ariano, Perinotti, 2011).

Now, the strange thing about this analysis is that even if you don't believe in a concept such as the wave function collapse, the end result will still be the same, since a non-superposed photon will be containing the same information, just as if it had collapsed from a superposed state, and will be subject to the same S-E law. The photon *has* to pass on its energy and especially its *information* to a retinal molecule every time, which is the commencement of this process, subject to the quantum detection efficiencies. One could perhaps more simply say that "Schrödinger meets Stark-Einstein", and forget all this accompanying analysis!

**4. Applicability of the von Neumann collapse postulate?**

We now address the issue of how or whether von Neumann's two-process collapse postulate might represent the best fit in the above scenario as compared to all the other interpretive postulates (von Neumann, 1955). The $1^{st}$ process states that light or the measured quantum system *S* interacts with a macroscopic measuring apparatus *M* for some physical quantity *Q*, with the interaction governed by the linear deterministic Schrödinger equation. At first glance this appears to come into conflict with this analysis, in that in our case the measured quantum system *S* interacts with a microscopic measuring apparatus *M*, the retinal molecule. Except that it is part of a rod cell which is 2 μm in dia. x 100 μm in length, and which is also a part of the retina, which is macroscopic. One can then say that the microscopic retinal molecule can be considered part of the macroscopic retina via entanglement and von Neumann is correct.

The 2nd process states that after this first stage of the measurement terminates, and one has a linear combination of products which are called entangled states, a 2nd nonlinear indeterminate process takes place, the collapse of the wave function. Once again this appears to conflict with our proposal, in that a determinate nonlinear process has taken place and that we have gone from deterministic linear states to a deterministic nonlinear state. And, that it was a process governed by the retinal molecule operating in accordance with the S-E law.

It should be noted here that we are still at the quantum level and we know exactly where we are at all times in this evolution from microscopic to macroscopic. We will know at what point superpositions break down and definite outcomes appear even at this level. As has been pointed out previously, this must be a very simple and generic process which is applicable to more than $1.3 \times 10^6$ other living entities, i.e. in the visual receptors of all three phyla which possess eyes: mollusks, arthropods and vertebrates (Sugihara et al, 2002; Fernald, 2004; Thaheld, 2009).

**5. The Internalist approach**

This analysis embraces what is known as the *internalist* stance proposed by Matsuno (Matsuno, 1989, 1996). This means that the material act of distinguishing between before and after physical events, whatever they are, to be most fundamental, irreducible and even ubiquitous inside this empirical world. The linear approach, no matter how cherished by the majority, would remain secondary at best. Once one accepts this stance, nonlinearity intrinsic to the internal act of making distinctions, would turn out to be the rule rather than the exception, which will represent a new doctrine (Matsuno, 1989, 1996). And, that although the Schrödinger equation of the wave function is linear, the preparation of the boundary conditions is nonlinear because any material bodies are involved in internal measurement (Conrad and Matsuno, 1990; Matsuno, 2003). He also feels that the idea of *naturalization* of the collapse (when taken in league with the *naturalization* of the Born rule) underlies the whole issue of quantum mechanics (Thaheld, 2009).

**6. Discussion**

Our world is awash in massive amounts of *classical environmental information* prepared by Nature, which represents *maximal* deterministic knowledge for us about the system's preparation. With Nature, in this specific instance, there is always probability 1 as far as we are concerned, which implies either *pure* states or *mixed* states as part of a *pure* state (Chiribella, D'Ariano, Perinotti, 2011). When a photon interacts with this *information* it instantly acquires it and reduces it to a quantum level, where it is converted to a wave function while still retaining its inherent *extrinsic* classicality. The wave function will then commence evolving in a linear superposition of deterministic states, as per the Schrödinger equation, up to the point where it comes into contact with the retinal molecule and the S-E law.

When the photon is absorbed by the retinal molecule the linear evolution becomes nonlinear and the *information* and energy being carried by the photon is now passed on to a π electron in the retinal molecule which then jumps to a π* electron orbit commencing the process finally leading to phototransduction. You will note that this is a *reversible* process as regards to the *information*, with the photon first acquiring this *information* from Nature and reducing it down to the quantum level, then giving it up or transferring it to the retinal molecule in a *reversible* fashion, where it is then amplified back up to the classical level by the rod cell.

The reviewer has asked if there is any decisive scheme for distinguishing between *intrinsic* quantum information and *extrinsic* classical information as it relates to the photon? It is possible to distinguish between them since all of the *intrinsic* information is an inherent component of the photon makeup at the quantum level while the *extrinsic* information starts out at the classical level, is then acquired by the photon and reduced to the quantum level as an addition, and then disposed of by transference to a retinal molecule, where a process of amplification commences bringing it back to the classical level. The energy of the photon is passed on to a π electron simultaneously with the information. The *intrinsic* information is of a limited and given known amounts while the *extrinsic* information is of an infinite and variable amount representing maximum *information* or *knowledge* relating to its preparation, acquisition and disposition.

I am further indebted to the reviewer of my manuscript for bringing to my attention the subject of QBism or Quantum Bayesianism and its possible relationship to this analysis. Now, after reading several papers on the subject, I suddenly realize that I have been a 'closet QBist' all along with a probability 1 (Fuchs and Schack, 2013; Mermin, 2014)! Especially since I had previously raised the issue of what I referred to as the 'naturalization of the Born rule' or 100% probability, and also the role of determinism (Thaheld, 2009).

Some people promote the idea of QBism as paying attention to the *user* of quantum mechanics. The reviewer asked if such a *user* can serve as the interface agent between the deterministic unitary quantum evolution and the deterministic classicality in the light of the present work? My answer to this is a resounding YES! That in this specific biological setting the *user* represents a most indispensable and critical interface. And, when I say '*user*', I do not have the hubris to apply the term to just homo sapiens alone but, to all of the previously mentioned phyla, of which there are over $1.3 \times 10^6$ in our world (Thaheld, 2009)! Amazingly, the *user* also represents the measurement device which is normally an external macroscopic entity. Without the *user* this paper would lose most of its meaning and we would probably not exist to even be discussing this matter. It appears that the interaction between the retinal molecule and the S-E law may have played a critical role in evolution and the survival of species.

To better illustrate, just close your eyes and all the photon energy and information will impinge upon your eyelids and is forever lost! Open them up again and instantaneously this energy and information is back to performing the most critical function in the universe as far as we are concerned. And, you have just performed a fundamental quantum experiment which will always have a probability 1!

7. **Conclusion**

In conclusion it is important to stress here that no linear superposed photon states can get past the S-E law, as it is applied to the retinal molecule in this specific instance. That it will always act as a boundary, and any linear visual photon states will always be converted or reduced to a nonlinear non-superposed state, and this will occur in the same place each time for any one of the retinal molecules at the quantum level. The only thing that gets past is energy and the original *classical environmental information*, still in the same deterministic state as when it first started out, most importantly as prepared by Nature*,* and waiting to get amplified back from the quantum to the classical world.

This means that we will now be able to accurately determine the Heisenberg 'cut' between these 2 worlds and there will be no need to invoke the von Neumann 'chain'. And, if we repeat the same or any *classical environmental* measurement time after time, it will always lead to definite and similar outcomes for any observers, and the distribution of these outcomes can be given by a modification referred to as the *naturalization* of the Born rule with probability 1. I will leave it up to the reader whether she wants to still call this a measurement or now refer to this process as an observation of *environmental information*.

In the near future it may be possible to subject this proposal to empirical tests, as has been outlined before, especially now in light of recent experiments utilizing single photons to stimulate retinal rod cells (Phan et al, 2013; Thaheld, 2003). This could in turn lead to an experiment to determine just how much *classical environmental information* a photon can carry, and whether this is of a sub-quantum nature, since the photon itself is already a quantum unit. The answer may reside in an examination of the extremely small visual receptors of the smallest of the three phyla, which apparently enable them to visualize the world much as we observe it. One direction might involve the use of mice retinal knock-out genes.

It appears that the Schrödinger linear equation will have to be modified to include these nonlinear changes. The quantum definition of physical units such as the second, the meter, the Ohm and electrical charge among others, heavily rely on the validity of basic dynamical equations like the Schrödinger equation. If this was to be modified then some definitions of physical units may also be affected (Amelino-Camelia et al, 2005).

While I did not mention the cone cells in the retina, this modification of the S-E law would apply equally to them as it has to the rod cells.

The author also finds it most interesting that we are reaching back over 100 years to find a possible solution within the S-E law to a problem that so interested and vexed Einstein and many others. A solution which will once again hopefully enable us to restore determinism to its proper role after having been missing from the scene for over 2,000 years (Heisenberg, 2011).

Finally, I am sure that some of you have already recognized at least two problems hanging over this analysis, the first one regarding whether the wave function is real or just a mathematical construct and the second one of pre-determination and its implication for the uncertainty principle and superluminal communication (Gisin, 2009)!

The author had considered ending this discourse at this point, as he felt that he may have already stretched quantum credulity to the breaking point and beyond. However, the thought occurred that there may be something here that might be applicable to the measurement problem from the *inanimate* perspective. I.e., can this concept of determinism be carried beyond the *animate* biological realm into the *inanimate* quantum mechanical world to some degree? It is in this spirit that I am simply raising the question as to whether it would be possible to provide one of two entangled photons, with some type of *additional* simple information, over and above its polarization states (which I am sure we would all agree represent a paucity of information), while the photons are in transit to their respective detectors, without causing a collapse of the wave function at the point of insertion of the information? It has been suggested to the author that this might be accomplished by the use of tiny prisms that reflect light in different ways, so as to obtain a colored image (say an abstract one at the beginning) on the other side (Pizzi, 2015).

**Acknowledgement**

I wish to thank the reviewer for pointing out certain areas which needed clarification and also the many people who kindly let me pose a range of difficult and controversial questions over the years. To Eric Thaheld who provided the impetus for me to finally undertake this analysis. And, to Brenda Hall for proof reading and correcting the revisions.

Amelino-Camelia, G., Lammerzahl, C., Macias, A., Muller, H. 2005. The search for quantum gravity signals. arXiv:gr-qc/0501053.

Baylor, D.A., 1996. How photons start vision. Proc. Natl. Acad. Sci. USA. 93, 560-565.

Chiribella, G., D'Ariano, G.M., Perinotti, P., 2010. Probabilistic theories with purification. Phys. Rev. A 81, 062348.

Chiribella, G., D'Ariano, G.M., Perinotti, P., 2011. Informational derivation of Quantum Theory. Phys. Rev. A 84, 012311.

Conrad, M., Matsuno, K., 1990. The boundary condition paradox: A limit to the universality of differential equations. Appl. Math. Comput. 37, 67-74.

Cox, A., Kemp, T.J., 1971. Introductory Photochemistry. McGraw-Hill, London.

Fernald, R.D., 2004. Evolving eyes. Int. J. Dev. Biol. 48, 701-705.

Fuchs, C.A., Schack, R., 2013. Quantum-Bayesian coherence. Rev. Mod. Phys. 85, 1693-1715.

Gisin, N., 2009. Quantum nonlocality: How does Nature perform the trick? arXiv:0912.1475.

Hardy, L., 2001. Quantum theory from five reasonable axioms. arXiv:quant-ph/0101012 v4.

Heisenberg, W., 2011. Is a deterministic completion of quantum mechanics possible? English translation of an unpublished manuscript by E. Crull and G. Bacciagaluppi, http://philsci-archive.pitt.edu/8590/

Leggett, A.J., 2005. The quantum "measurement problem". Science 307, 871-872.

Matsuno, K., 1989. Protobiology: Physical Basis of Biology. CRC Press, Boca Raton. Chap. 2, Internal Measurement. 31-55.

Matsuno, K., 1996. Internalist stance and the physics of information. BioSystems 38, 111-118.

Matsuno, K., 2003. Quantum mechanics in first, second and third person descriptions. BioSystems 68, 1-12.

Mazzolini, M., Facchetti, G., Andolfi, L., et al, 2015. The phototransduction machinery in the rod outer segment has a strong efficacy gradient. Proc. Natl. Acad. Sci. USA. May 19; 112 (20): E2715-24. doi:10.1073/pnas.1423162112.

Mermin, N.D., 2014. Why QBism is not the Copenhagen interpretation and what John Bell might have thought of it. arXiv:1409.2454.
10